\documentclass[12pt]{iopart}
\usepackage{iopams}
\usepackage[T1]{fontenc}
\usepackage[latin9]{inputenc}
\usepackage{varioref}
\usepackage{units}
\usepackage{graphicx}
\usepackage{esint}
\usepackage{caption}
\usepackage{subcaption}
\usepackage{siunitx}
\usepackage{hyperref}
\usepackage{lineno}

\begin{document}

\title{Study of the band-gap energy of radiation-damaged silicon}

\author{R.~Klanner$^1$, S.~Martens$^1$, J.~Schwandt$^1$ and  A.~Vauth$^1$}

\address{$^1$ Institute for Experimental Physics, University of Hamburg, Luruper Chaussee 149, 22761, Hamburg, Germany.}

\ead{Robert.Klanner@desy.de}
\vspace{10pt}
\begin{indented}
\item[]March 2022
\end{indented}


\begin{abstract}

 The transmission of silicon crystals irradiated by 24\,GeV/c protons and reactor neutrons has been measured for photon energies, $E_\gamma$, between 0.95 and 1.3\,eV.
 From the transmission data the absorption coefficient $\alpha$ is calculated, and from $\alpha (E_\gamma)$ the fluence dependence of the band-gap energy, $E_\mathit{gap}$, and the energy of transverse optical phonons, $E_\mathit{ph}$, determined.
 It is found that within the experimental uncertainties of about 1\,meV neither $E_\mathit{gap}$ nor $E_\mathit{ph}$ depend on fluence up to the maximum fluence of $1 \times 10^{17}\mathrm{cm}^{-2}$ of the measurements.
 The value of $E_\mathit{gap}$ agrees within about 1\,meV with the generally accepted value, if an exciton-binding energy of 15\,meV is assumed.
 A similar agreement is found for $E_\mathit{ph}$.
 For the extraction of $E_\mathit{gap}$ and $E_\mathit{ph}$ the second derivative of $\sqrt{ \alpha (E_\gamma) }$ smoothed with a Gaussian kernel has been used.

\end{abstract}

%
\vspace{2pc}

\section{Introduction}
 \label{sect:Introduction}
 The band-gap energy, $E_\mathit{gap}$, is a fundamental parameter characterising semi-conductors.
 Its knowledge is required for a quantitative understanding of properties like
 intrinsic charge carrier density, Fermi level, dark current, emission probabilities and many others.
 Whereas $E_\mathit{gap}$ of crystalline silicon has been studied in detail as a function of temperature and doping, there are hardly any investigations on its dependence on radiation damage.
 As radiation damage produces point and cluster defects, the situation is quite complex and there are no reliable estimates on the expected change of $E_\mathit{gap}$ as a function of particle type and fluence.

 Silicon is an indirect semi-conductor and to excite an electron from the valence to the conduction band by a photon with an energy, $E_\gamma$, close to $E_\mathit{gap}$, requires a phonon to satisfy energy and momentum conservation.
 It is expected that optical-transverse phonons with an energy $E_\mathit{ph}$ dominate, and that the light absorption $\alpha (E_\gamma)$ can be described by\,\cite{Yu:2010}


 \begin{small}
  \begin{equation}\label{eq:alpha}
   \hspace{-2.2cm}
    \alpha(E_\gamma, T) \propto
    \frac{\left( \max (E_\gamma - E_\mathit{gap} + E_\mathit{ph} + E_\mathit{exc},0) \right)^2}{\exp (E_\mathit{ph}/k_B T)-1} +
    \frac{\left( \max (E_\gamma - E_\mathit{gap} - E_\mathit{ph} + E_\mathit{exc},0 \right))^2}{1 - \exp (-E_\mathit{ph}/k_B T)}
 \end{equation}
 \end{small}
 with the exciton binding energy, $E_\mathit{exc}$, the absolute temperature, $T$, and the Boltzmann constant, $k_B$.
 The first term describes the phonon absorption, and the second one the phonon emission.
 As pointed out in \cite{Tsai:2018}, there are eight routes of second-order transitions in an indirect band-gap semiconductor between a state in the conduction band and the valence band via an intermediate state, and Eq.\,\ref{eq:alpha} is only approximate.
 In addition, there can be contributions to $\alpha $ from states in the band gap, which is expected to be of particular relevance in the case of radiation damage.

 In \cite{Scharf:2020} phosphorous-doped silicon crystals with 3.5~k$\Omega \cdot$cm resistivity and \SI{280}{\um} thickness were irradiated with 24 GeV/c protons up to a 1~MeV neutron-equivalent fluence of $\Phi _\mathit{eq} = 8.6 \times 10^{15}$~cm$^{-2}$, and the light transmission measured for wavelenghts between 0.95 and \SI{1.3}{\um}.
 In the analysis only the phonon-emission term has been taken into account, and a linear narrowing of $E_\mathit{gap}$ with a slope of d$E_\mathit{gap} / \mathrm{d} \Phi _\mathit{eq} \approx -\, 6\times 10^{-16}~\mathrm{meV}\cdot \mathrm{cm}^2$ is reported.
 In the conclusions it is stated:
  \emph{For the band-gap narrowing due to radiation damage further studies are required to verify the results presented in this paper and refine the analysis methods.}
 This is the topic of the present paper.

 The next section gives an overview over the samples and the optical transmission measurements.
 In addition to the samples of \cite{Scharf:2020}, silicon crystals of 3\,mm thickness irradiated by reactor neutrons up to fluences of $1 \times 10^{17}$\,cm$^{-2}$ have been used, and isochronal annealing up to a temperature of 210\,$^\circ $C has been performed.
 An analysis of the absorption data for non-irradiated silicon summarized by Green in\,\cite{Green:2021} follows.
 It is found that Eq.\,\ref{eq:alpha} does not describe the photon absorption of non-irradiated silicon close to $E_\mathit{gap}$.
 Therefore, another method, called \emph{smoothed derivative} method in the following, is proposed: $E_\mathit{gap}$ and $E_\mathit{ph}$ are determined from the maxima of the second derivative of $\sqrt{\alpha (E_\gamma)}$ smoothed with a Gaussian kernel.
 This method is then used to analyse the data from the irradiated silicon samples, and the $\Phi _\mathit{eq}$ dependence of $E_\mathit{gap}$, $E_\mathit{ph}$
 is presented.
 An attempt to extract the band-gap energy inside damage clusters follows, and the main results are summarised in the final section.

 \section{Samples and light-transmission measurements}
  \label{sect:Samples}

 For the study phosphorous-doped float-zone silicon crystals with approximately 3.5\,k$\Omega \cdot$cm resistivity were used.
 The samples, already presented in \cite{Scharf:2020}, with a thickness of \SI{280}{\um}, were irradiated by 24\,GeV/c protons to $\Phi _\mathit{eq} = $ (2.4, 4.9, 6.1, 8.6)$\times 10^{15}\,\mathrm{cm}^{-2}$\,\cite{CERN:Irrad}, where a hardness factor $\kappa = 0.62$\,\cite{Moll:2002} was used.
 The new samples have a thickness of 3\,mm.
 They were irradiated by reactor neutrons \cite{IJS:Irrad} to fluences of $\Phi _\mathit{eq} = $ (1, 5, 10, 30, 50, 100)$\times 10^{15}\,\mathrm{cm}^{-2}$.
 The irradiations were performed approximately at room temperature, however the precise value is not known.
 The estimated uncertainties of $\Phi _\mathit{eq}$ are 10\,\%.

\begin{figure}[!ht]
   \centering
   \begin{subfigure}[a]{0.48\textwidth}
    \includegraphics[width=\textwidth]{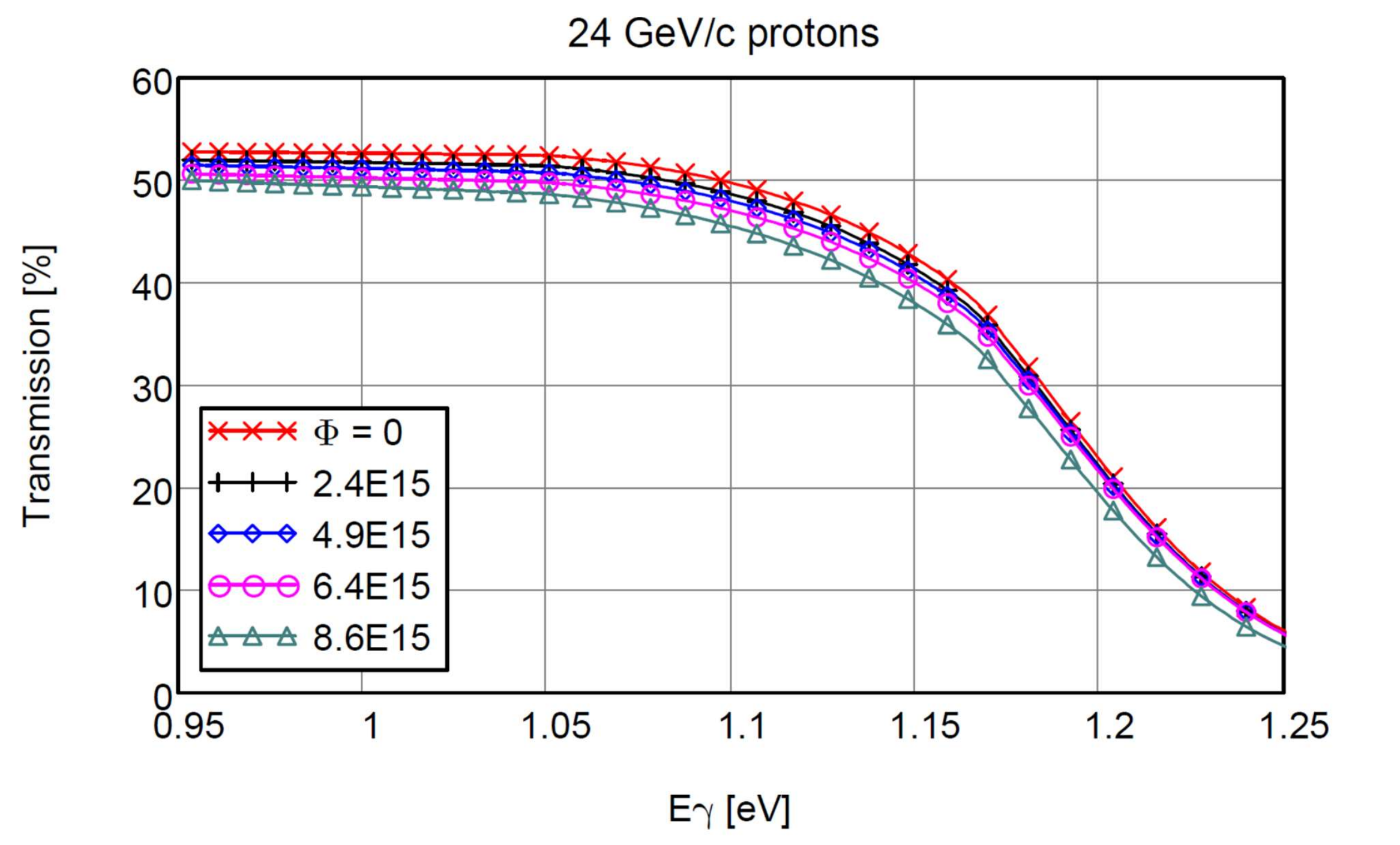}
    \caption{ }
    \label{fig:FigTr-protons}
   \end{subfigure}%
    ~
   \begin{subfigure}[a]{0.5\textwidth}
    \includegraphics[width=\textwidth]{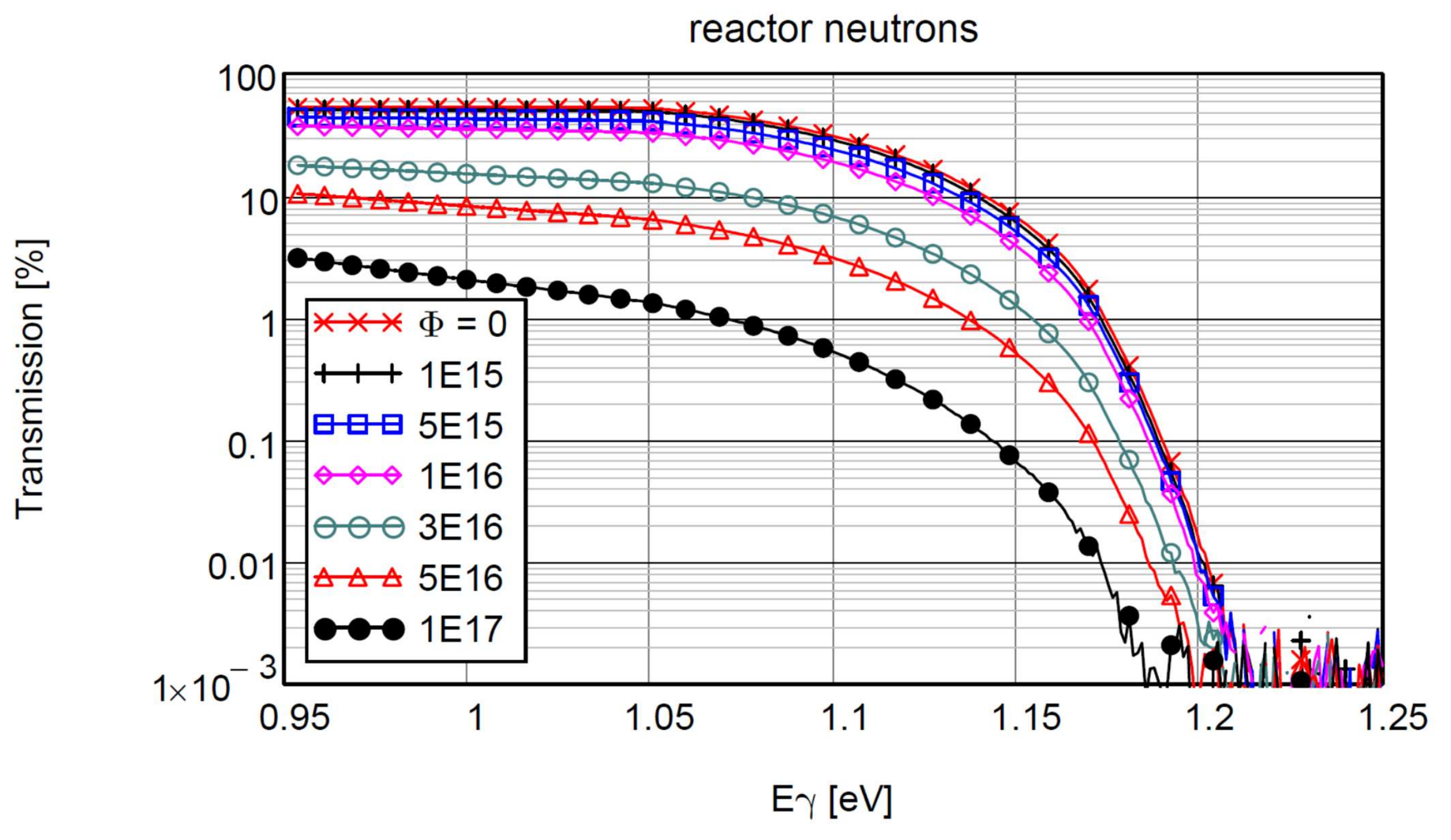}
    \caption{ }
    \label{fig:FigTr-neutrons}
   \end{subfigure}%
   \caption{Measured transmission as a function of $E_\gamma$ for
   (a) the silicon samples of thickness $d = \SI{280}{\um}$ irradiated with 24 GeV/c protons, and
   (b) for the samples of $d = 3$\,mm irradiated with reactor neutrons. For $E_\gamma \gtrsim 1.2$\,eV the transmission of the 3\,mm samples is below 0.01\,\%, and the measurements are dominated by noise.
   In this and in other figures markers are used to distinguish the different curves.
   The distance in $E_\gamma $ of the individual measurement points is about 1\,meV. }
  \label{fig:FigTr}
 \end{figure}

 For the transmission measurements an Agilent CARY 5000 UV-VIS-NIR \cite{Agilent} was used.
 The wavelength range for the \SI{280}{\um} samples was 0.95 to \SI{1.35}{\um}, and for the 3\,mm samples 0.95 to \SI{2}{\um}.
 Fig.\,\ref{fig:FigTr} shows the transmission measured at a temperature of about 295\,K as a function of the photon energy, $E_\gamma $\,[eV] = 1.24$/\lambda \,[\si{\um}]$.

 For the isochronal annealing the 3\,mm samples were heated for 15 min to temperatures of (80, 100, 120, 150, 180, 210)\,$^\circ $C.

 From the measured transmission, $Tr(\lambda)$, the absorption coefficient, $\alpha (\lambda)$, is obtained using the standard formulae \cite{Scharf:2020}:

 \begin{equation}\label{eq:a-data}
  \alpha(\lambda) = \frac{1}{d} \cdot \ln\Bigg(\frac{\mathit{Tra}(\lambda) ^2+\sqrt{\mathit{Tra}(\lambda)^4 + 4 \cdot \mathit{Ref}(\lambda)^2 \cdot \mathit{Tr}(\lambda) ^2} } {2 \cdot \mathit{Tr}(\lambda)} \Bigg)
 \end{equation}
 with $\mathit{Ref}(\lambda) = \big(n(\lambda) - 1 \big)^2 / \big(n(\lambda) + 1 \big)^2 $ and
 $\mathit{Tra}(\lambda) = 1 - \mathit{Ref}(\lambda)$.
 The thickness of the sample is $d$, $\mathit{Ref}(\lambda)$ is the reflection of a single air-silicon interface, $\mathit{Tra}(\lambda)$ the corresponding transmission, and $n(\lambda)$ the index of refraction of silicon taken from \cite{Green:2021}, which is assumed not depend on the irradiation fluence.

 \section{Analysis of the absorption data for non-irradiated silicon}
  \label{sect:Green}

 In this section, it is investigated if Eq.\,\ref{eq:alpha} is able to describe the absorption data of non- irradiated silicon from the literature. It is found that this is not the case and a different method for extracting $E_\mathit{gap}$ and $E_\mathit{ph}$ is proposed.

 Inspecting Eq.\,\ref{eq:alpha} reveals that $\sqrt{\alpha (E_\gamma)}$ as function of $E_\gamma$ is expected to be
 zero for $E_\gamma < (E_\mathit{gap} - E_\mathit{ph} - E_\mathit{exc})$,
 a straight line for  $(E_\mathit{gap} - E_\mathit{ph} - E_\mathit{exc}) < E_\gamma < (E_\mathit{gap} + E_\mathit{ph} - E_\mathit{exc})$, where the second term is zero,
 followed by a second straight line for $E_\gamma \gg (E_\mathit{gap} + E_\mathit{ph}- E_\mathit{exc})$, where the second term dominates.
 For this reason it is customary to look at $\sqrt{\alpha (E_\gamma)}$.

  \begin{figure}[!ht]
   \centering
    \includegraphics[width=0.6\textwidth]{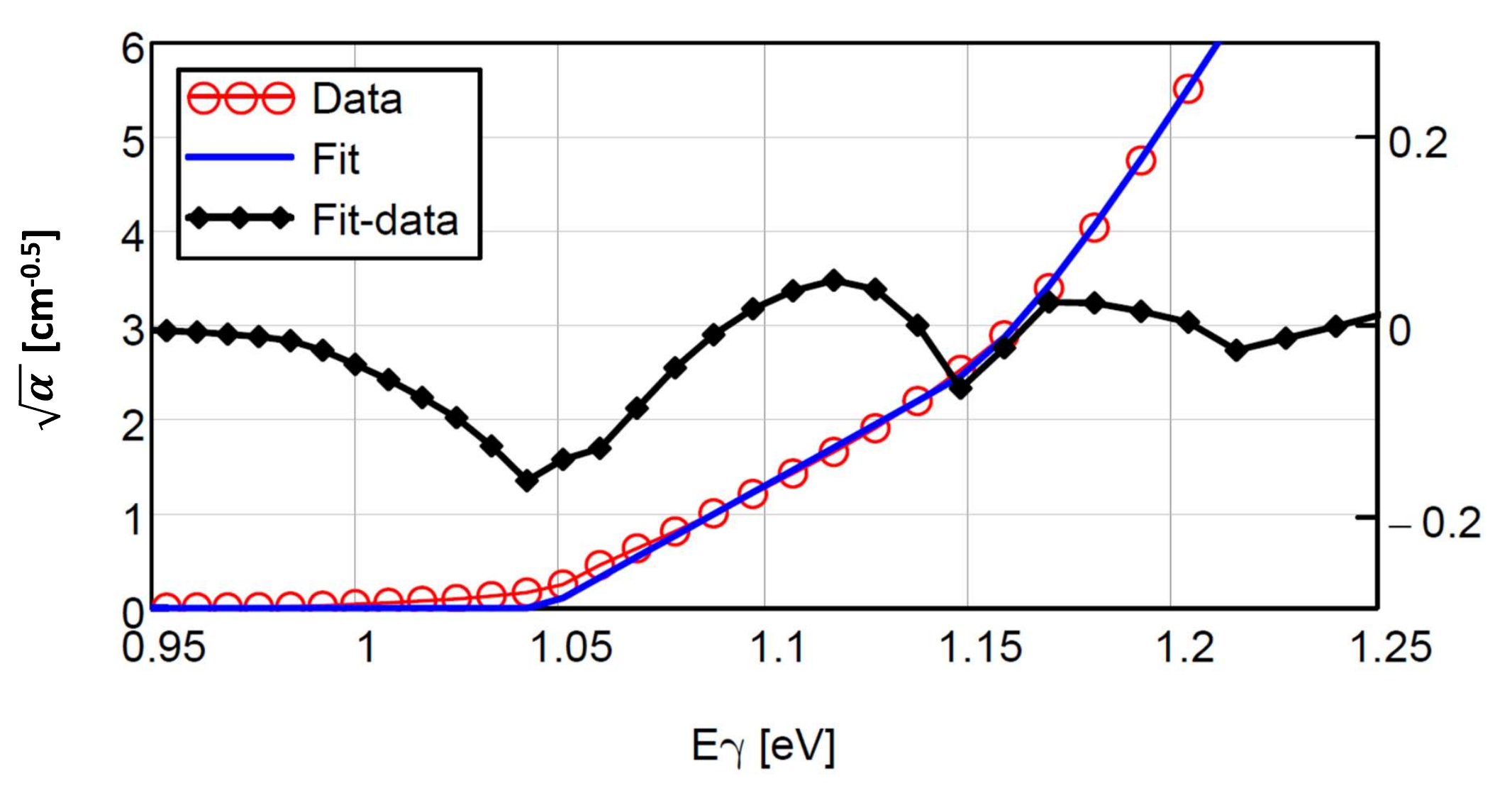}
   \caption{Comparison of $ \sqrt{ \alpha (E _\gamma ) } $ for the data of \cite{Green:2021} for non-irradiated silicon  to a fit using Eq.\,\ref{eq:alpha}.
   As shown by the difference fit -- data (scale on the right), the data are not described by Eq.\,\ref{eq:alpha}.
   The results of the fit are reported in Table\,\ref{tab:EgapGreen}.}
  \label{fig:Sqrt-aGreen}
 \end{figure}

 The $\sqrt{\alpha (E_\gamma)}$~data at 25\,$^\circ$C\,\cite{Green:2021} are fitted by Eq.\,\ref{eq:alpha}, with $E_\mathit{gap}$, $E_\mathit{ph}$, $T$, and the normalisation as free parameters.
 For $E_\mathit{exc}$ 15\,meV\,\cite{Shaklee:1970, Green:2013} is used.
 Table\,\ref{tab:EgapGreen} and Fig.\,\ref{fig:Sqrt-aGreen} present the results.
 Qualitatively, the data agree with the expectation of two straight lines, however the fit is not able to describe the data quantitatively.
 In addition, significant differences to the generally accepted values of $E_\mathit{gap} = 1.124 \,\pm 0.003 $\,eV\,\cite{Bludau:1974, Bensalem:2017}, and $E_\mathit{ph} = 58 \pm 1 $\,meV\,\cite{Chynoweth:1962, Shaklee:1970} are observed.
 The value obtained from the fit of $T$, which is constrained by the ratio of the two slopes of $\sqrt{\alpha (E_\gamma )}$, is quite compatible with the temperature 296.15\,K of the transmission data used in the analysis.

  \begin{table}[!ht]
   \centering
   \caption{Results for $E_\mathit{gap}$, $E_\mathit{ph}$ and $T$ for the fits of Eq.\,\ref{eq:alpha} and of the \emph{smoothed derivative} method to  $\sqrt{\alpha (E_\gamma ) }$ for non-irradiated silicon from \cite{Green:2021}, compared to values from literature.
   For the exciton binding energy 15\,meV has been assumed\,\cite{Schenk:1998}.
   The reference value $E_\mathit{gap}(T = 295\,\mathrm{K}) = 1.124$\,eV is obtained from\,\cite{Bensalem:2017} where an uncertainty of about $\pm\,3$\,meV is given, and the reference value of $E_\mathit{ph}$ is from\,\cite{Chynoweth:1962}.}
    \begin{tabular}{c|c|c|c}
    & $E_\mathit{gap}$ [eV] & $E_\mathit{ph}$ [meV] & $T$ [K] \\
    \hline \hline
    Literature & $1.124 \pm 0.003$ &  $58 \pm 1 $  & 296.15 \\
    \hline
   Fit to $\sqrt{\alpha}$                     & 1.1105 & 49.3 & 275.8 \\
   d$^2 \sqrt{\alpha}/\mathrm{d} E_\gamma ^2$ & 1.1212 & 59.1 & -- \\
    \hline
   \end{tabular}
  \label{tab:EgapGreen}
 \end{table}

\begin{figure}[!ht]
   \centering
    \includegraphics[width=0.6\textwidth]{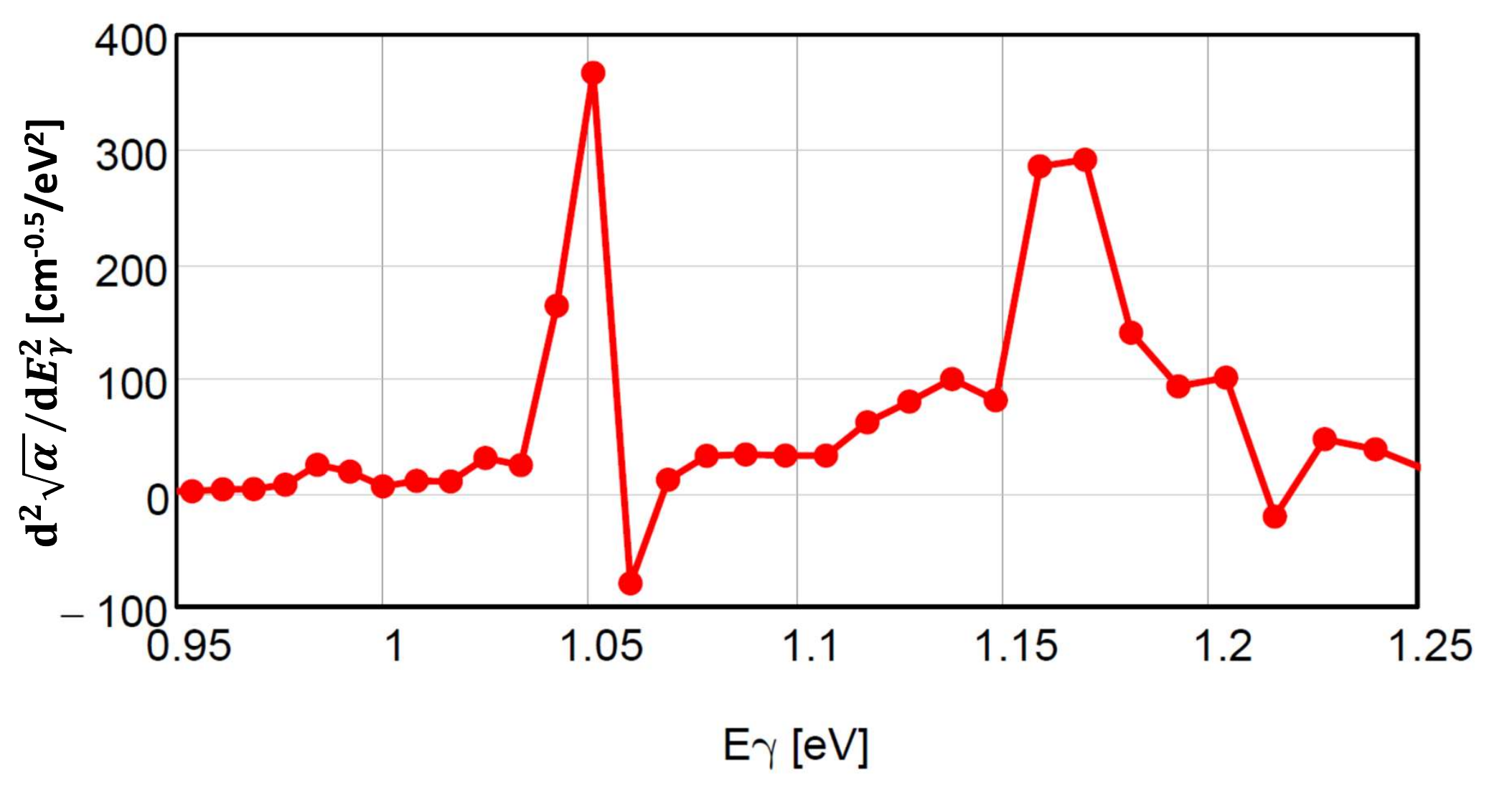}
   \caption{Smoothed second derivative d$^2 \sqrt {\alpha} / \mathrm{d} E _\gamma ^2 $ with $\sigma = 20$\,meV of the data of \cite{Green:2021} for non-irradiated silicon.
   The peaks are associated with the $E _\gamma$ values of phonon absorption and phonon emission, from which the values of $E_\mathit{gap}$ and $E_\mathit{ph}$, shown in Table\,\ref{tab:EgapGreen}, are derived.}
  \label{fig:SecDer-aGreen}
 \end{figure}

 The poor description of the data can be ascribed to additional contributions to the absorption or to scattering.
 Already for $E _\gamma $ below phonon absorption, the value of $\alpha$ is finite.
 Under the assumption that these additional contributions have smooth derivatives at the onsets of phonon absorption and  phonon emission, steps in the slope d$\sqrt{\alpha (E_\gamma )}/\mathrm{d}E_\gamma$ should hardly be affected, and maxima in d$^2\sqrt{\alpha (E_\gamma)}/\mathrm{d}E_\gamma ^2$ should appear at $E_\gamma = E_\mathit{gap} - E_\mathit{ph} - E_\mathit{exc}$ and $E_\gamma = E_\mathit{gap} + E_\mathit{ph} - E_\mathit{exc}$.
 Anticipating fluctuations for the measured $\alpha $ values, the following procedure, called \emph{smoothed derivative} method, is proposed:
  \begin{enumerate}
    \item Convolve $\sqrt{\alpha (E_\gamma )}$ with a Gaussian with variance $\sigma ^2$.
    \item Calculate numerically the second derivative of the convolved distribution.
    \item Associate $E_\gamma$ of the lower, dominant peak with $E_\mathit{gap} - E_\mathit{ph} - E_\mathit{exc} $, and the upper one with $E_\mathit{gap} + E_\mathit{ph} - E_\mathit{exc}$.
  \end{enumerate}
 Fig.\,\ref{fig:SecDer-aGreen} shows the smoothed second derivative using $\sigma = 20$\,meV.
 The results are not sensitive to the choice of $\sigma $.
 Narrow peaks are observed at $E_\gamma = 1.05$\,eV and at 1.17\,eV.
 Their precise values are obtained from the values of $E_\gamma $ at which the derivatives of d$^2\sqrt{\alpha(E_\gamma)}/\mathrm{d}E_\gamma ^2$ cross zero.
 Table\,\ref{tab:EgapGreen} presents the results.
 The value of $E_\mathit{ph}$ agrees with the literature values, and the value of $E_\mathit{gap}$ is 2.8\,meV lower, which can be considered satisfactory.
 For the data discussed in the next section, the differences are 1\,meV only, which is well within the uncertainties of the literature values of $E_\mathit{gap}$.

 \section{Analysis of the measured transmission data}
  \label{sect:Results}

 In this section, the \emph{smoothed derivative} method introduced in Sect.\,\ref{sect:Green} is used to extract the values of $E_\mathit{gap}$ and $E_\mathit{ph}$ from the measured transmission data.

\begin{figure}[!ht]
   \centering
   \begin{subfigure}[a]{0.5\textwidth}
    \includegraphics[width=\textwidth]{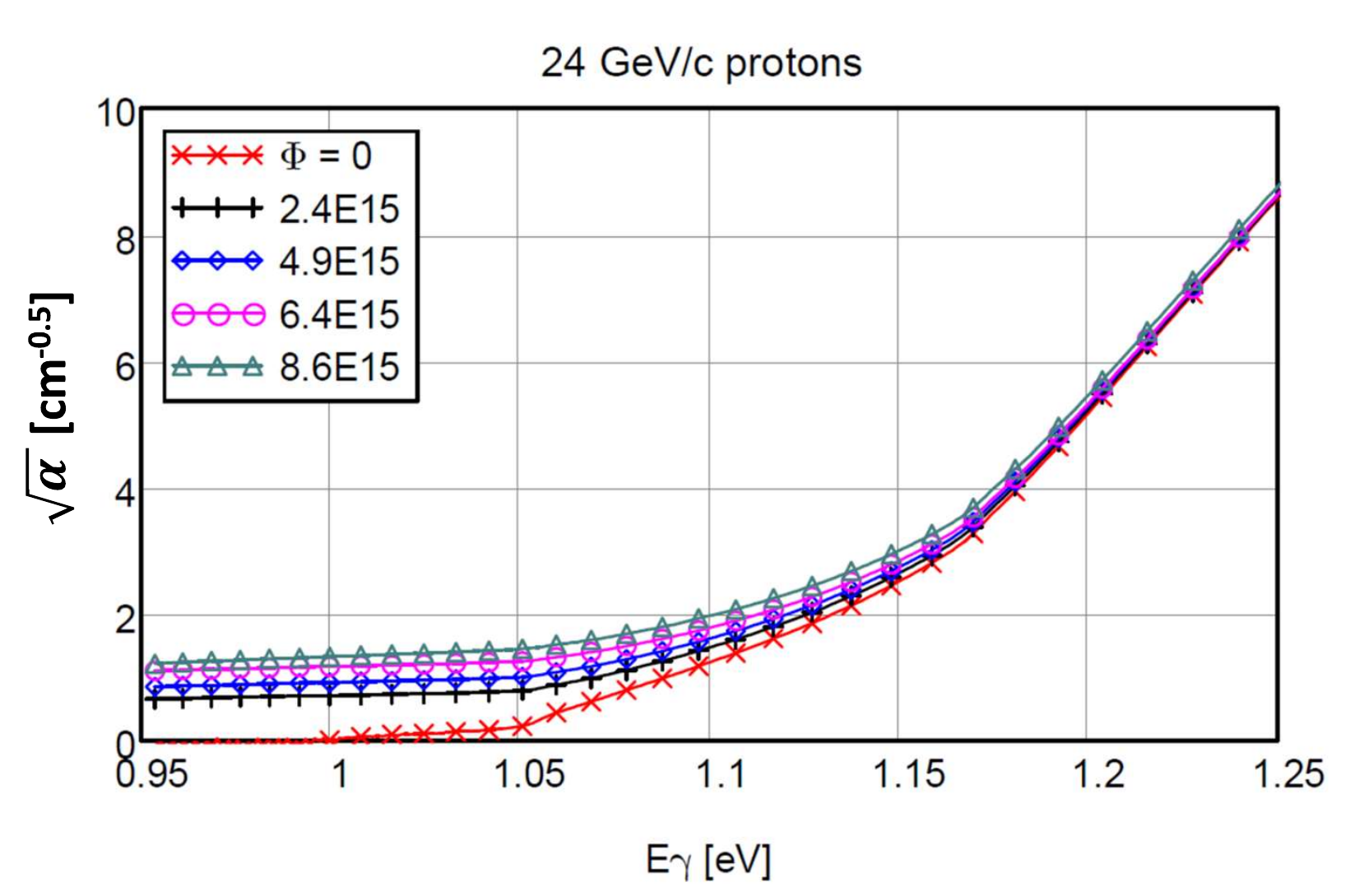}
    \caption{ }
    \label{fig:FigSqrta-protons}
   \end{subfigure}%
    ~
   \begin{subfigure}[a]{0.5\textwidth}
    \includegraphics[width=\textwidth]{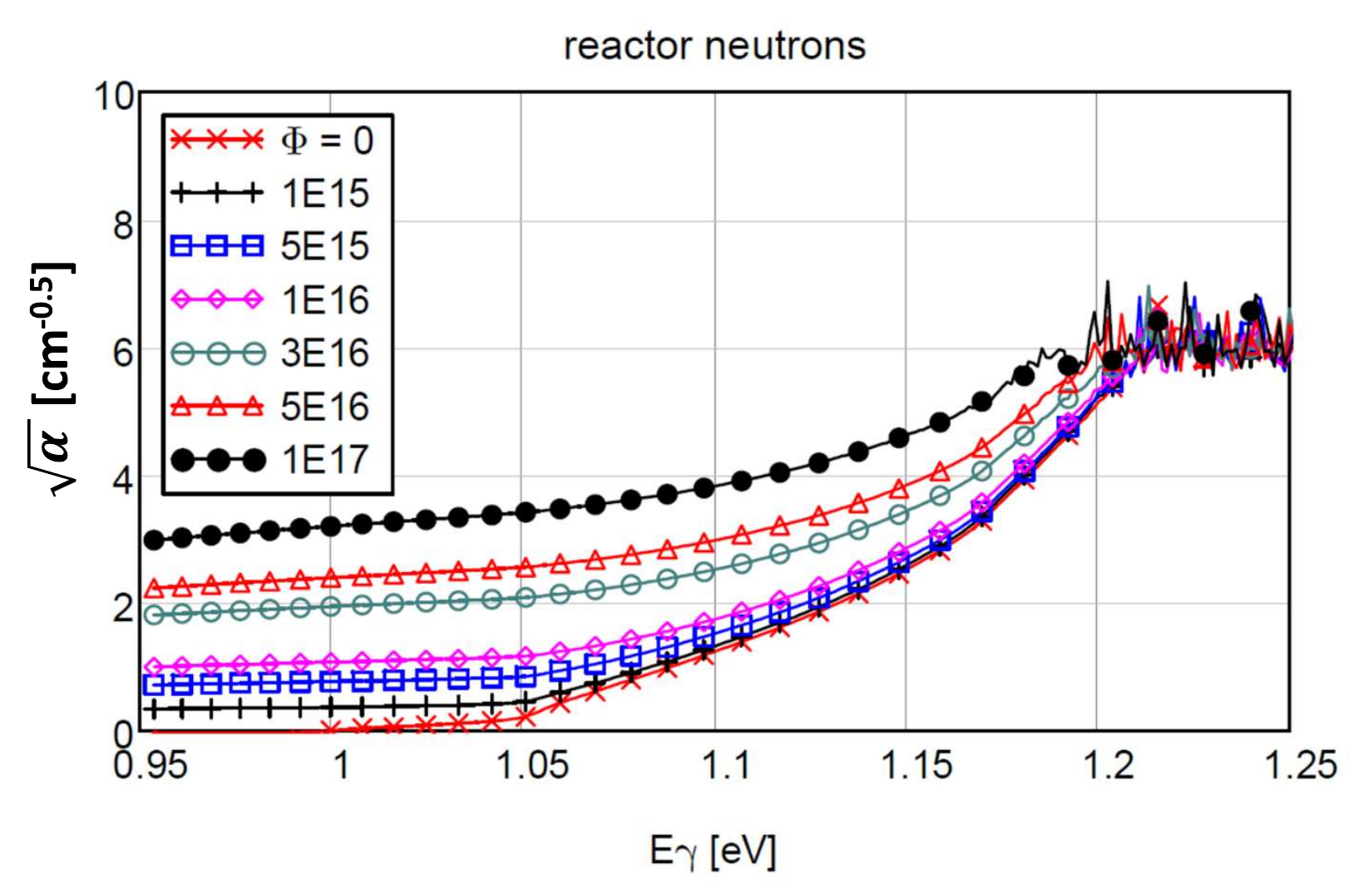}
    \caption{ }
    \label{fig:FigSqrta-neutrons}
   \end{subfigure}%
   \caption{$\sqrt{\alpha (E_\gamma ) }$ for
   (a) the silicon samples of thickness $d =\SI{280}{\um}$ irradiated with 24 GeV/c protons, and
   (b) for the samples of $d = 3$\,mm irradiated with reactor neutrons. For $E_\gamma \gtrsim 1.2$\,eV the transmission of the 3\,mm samples is below 0.01\,\%, and the $\alpha $ data are dominated by noise. }
  \label{fig:FigSqrta-irr}
 \end{figure}

 The data, $\sqrt{\alpha(E_\gamma)}$ for the different silicon thicknesses and $\Phi_\mathit{eq}$ values, are shown in Fig.\,\ref{fig:FigSqrta-irr}.
 It is noted that for the $d = 3$\,mm samples shown in Fig.\,\ref{fig:FigSqrta-neutrons}, the transmission for $E_\gamma \gtrsim 1.2$\,eV is below 0.01\,\% and the measurements are dominated by noise.
 It can also be seen that already for $E_\gamma < 1.05$\,eV, which is the photon energy required for the excitation of an electron from the valence to the conduction band by the absorption of a transverse optical phonon, the value of $\alpha $ is finite and increases with fluence.
 This increase is ascribed to the absorption by states in the band gap.
 As the $E_\gamma$ and $\Phi _\mathit{eq}$ dependence of this contribution is not known, a fit using Eq.\,\ref{eq:alpha} is problematic.

\begin{figure}[!ht]
   \centering
   \begin{subfigure}[a]{0.5\textwidth}
    \includegraphics[width=\textwidth]{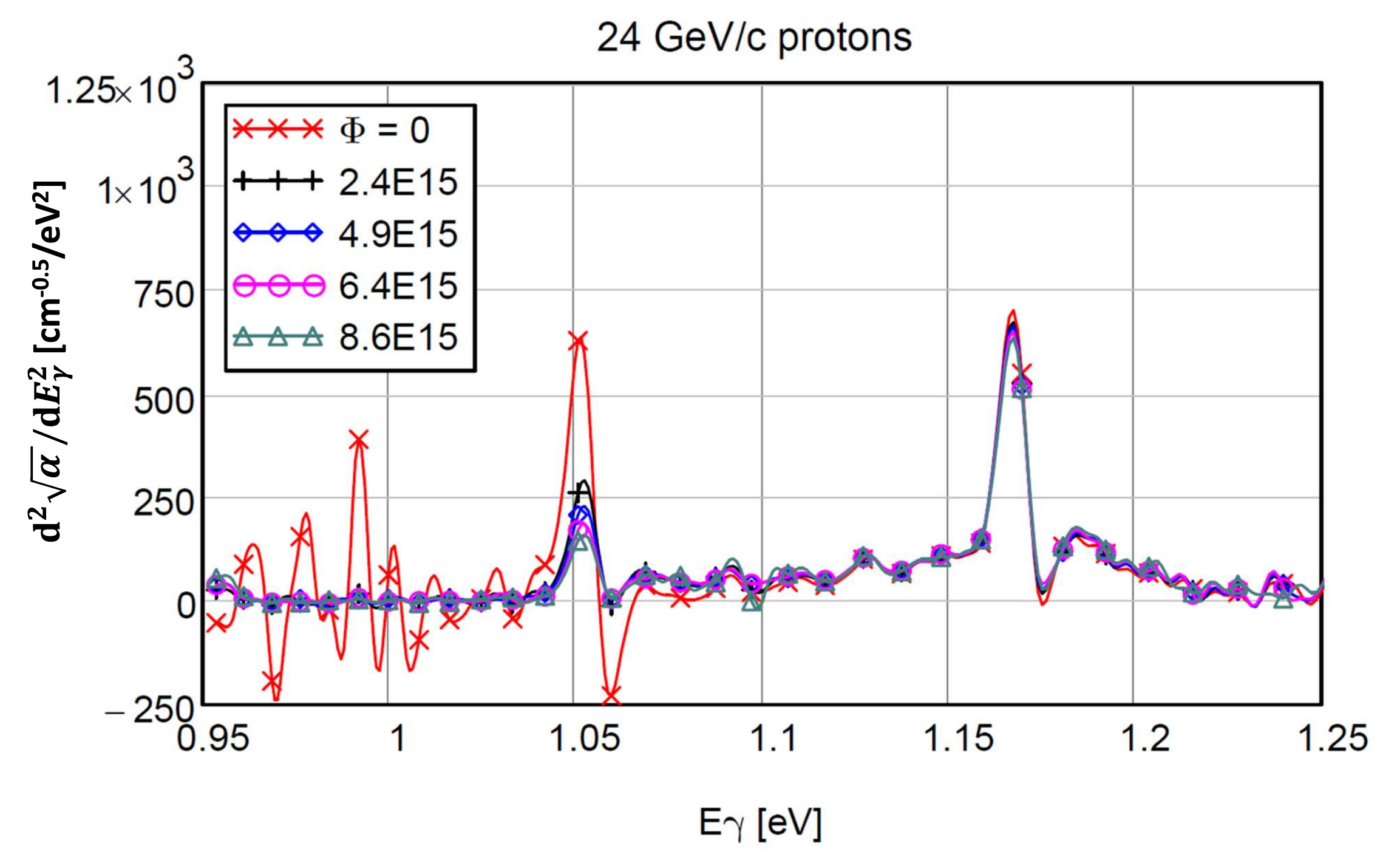}
    \caption{ }
    \label{fig:Figd2-protons}
   \end{subfigure}%
    ~
   \begin{subfigure}[a]{0.5\textwidth}
    \includegraphics[width=\textwidth]{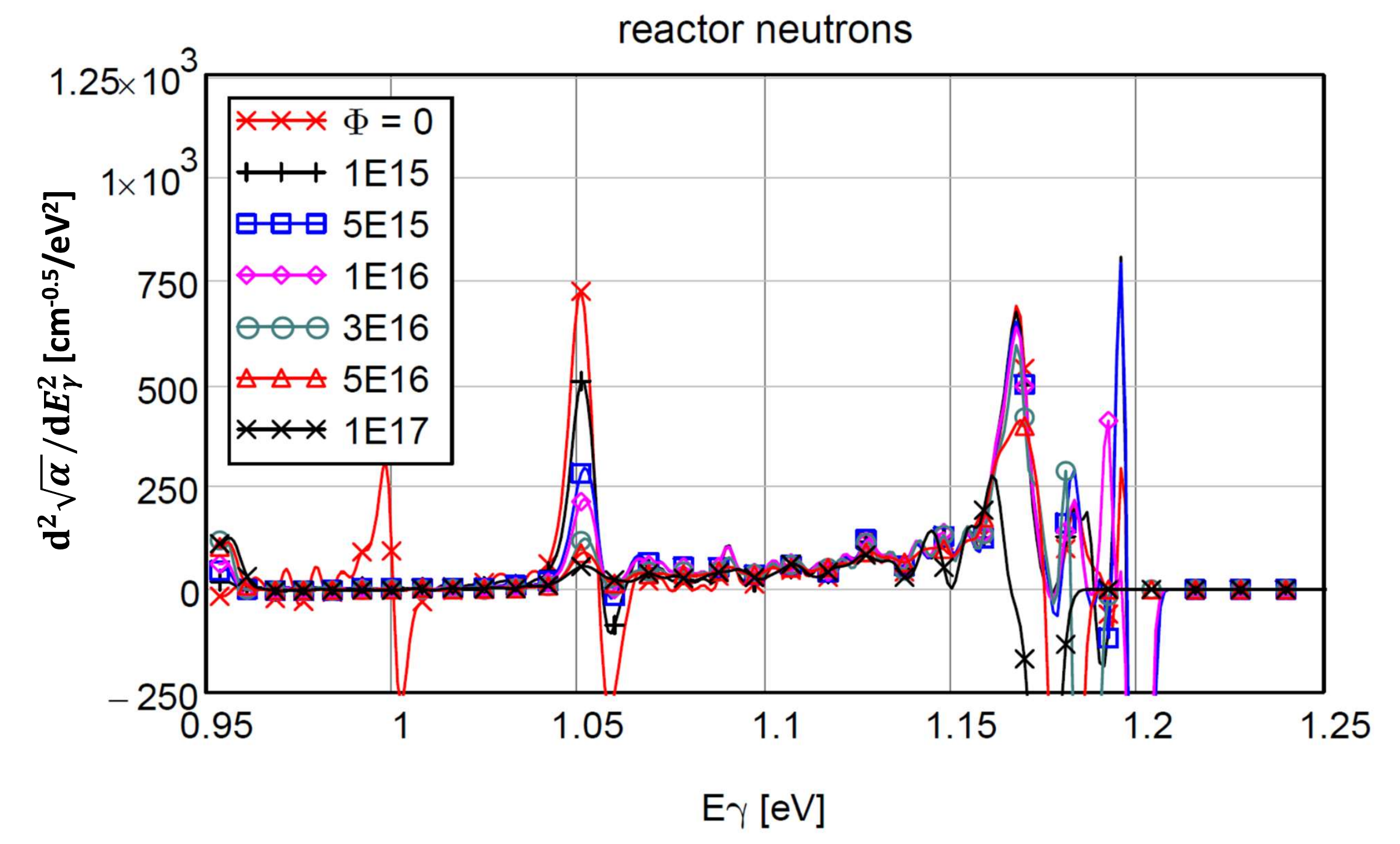}
    \caption{ }
    \label{fig:Figd2-neutrons}
   \end{subfigure}%
   \caption{Smoothed second derivative, d$^2\sqrt{\alpha (E_\gamma)}/\mathrm{d}E_\gamma ^2$, of the data shown in Fig.\,\ref{fig:FigSqrta-irr}.
    The peaks at 1.05\,eV are ascribed to phonon absorption, and the ones at 1.17\,eV to phonon emission.
   As noted in the caption of Fig.\,\ref{fig:FigTr}, the symbols on the lines are markers which are located at every tenth measured point.    }
  \label{fig:Figd2-data}
 \end{figure}

 Fig.\,\ref{fig:Figd2-data} shows the smoothed second derivative of the data of Fig.\,\ref{fig:FigSqrta-irr} using the value $\sigma = 20$\,meV for the Gauss convolution.
 Like for the data of non-irradiated silicon, sharp peaks are observed at $E_\gamma = 1.05$\,eV and 1.17\,eV.
 The results are not sensitive to the choice of $\sigma$.
 For the $\Phi _\mathit{eq} = 1 \times 10^{17}\,\mathrm{cm}^{-2}$\,data the noise in the transmission measurements for $E_\gamma \gtrsim 1.18$\,eV causes oscillations of the second derivative and the results for the 1.17\,eV peak can not be trusted.

\begin{figure}[!ht]
   \centering
   \begin{subfigure}[a]{0.5\textwidth}
    \includegraphics[width=\textwidth]{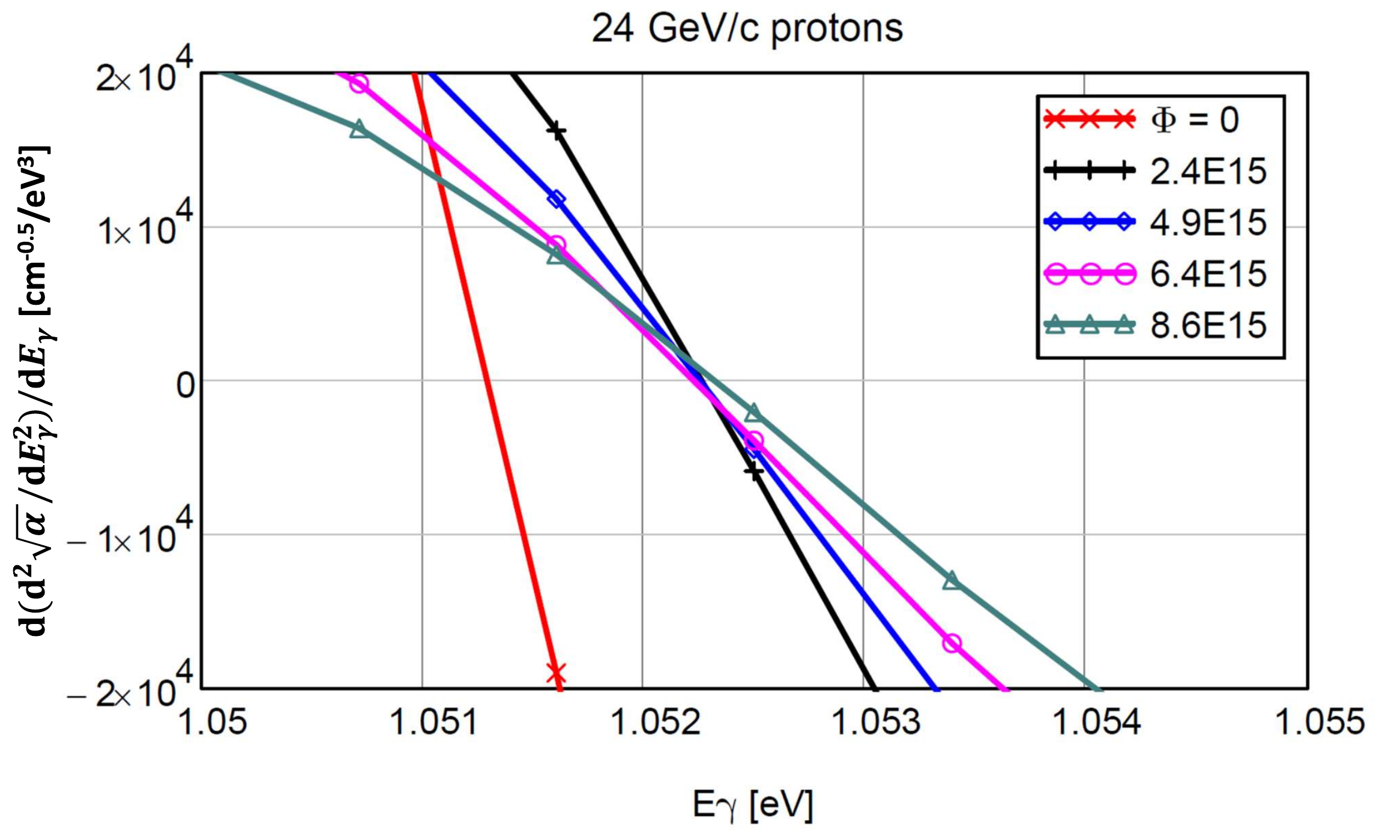}
    \caption{ }
    \label{fig:Figd3-protons-lo}
   \end{subfigure}%
    ~
   \begin{subfigure}[a]{0.5\textwidth}
    \includegraphics[width=\textwidth]{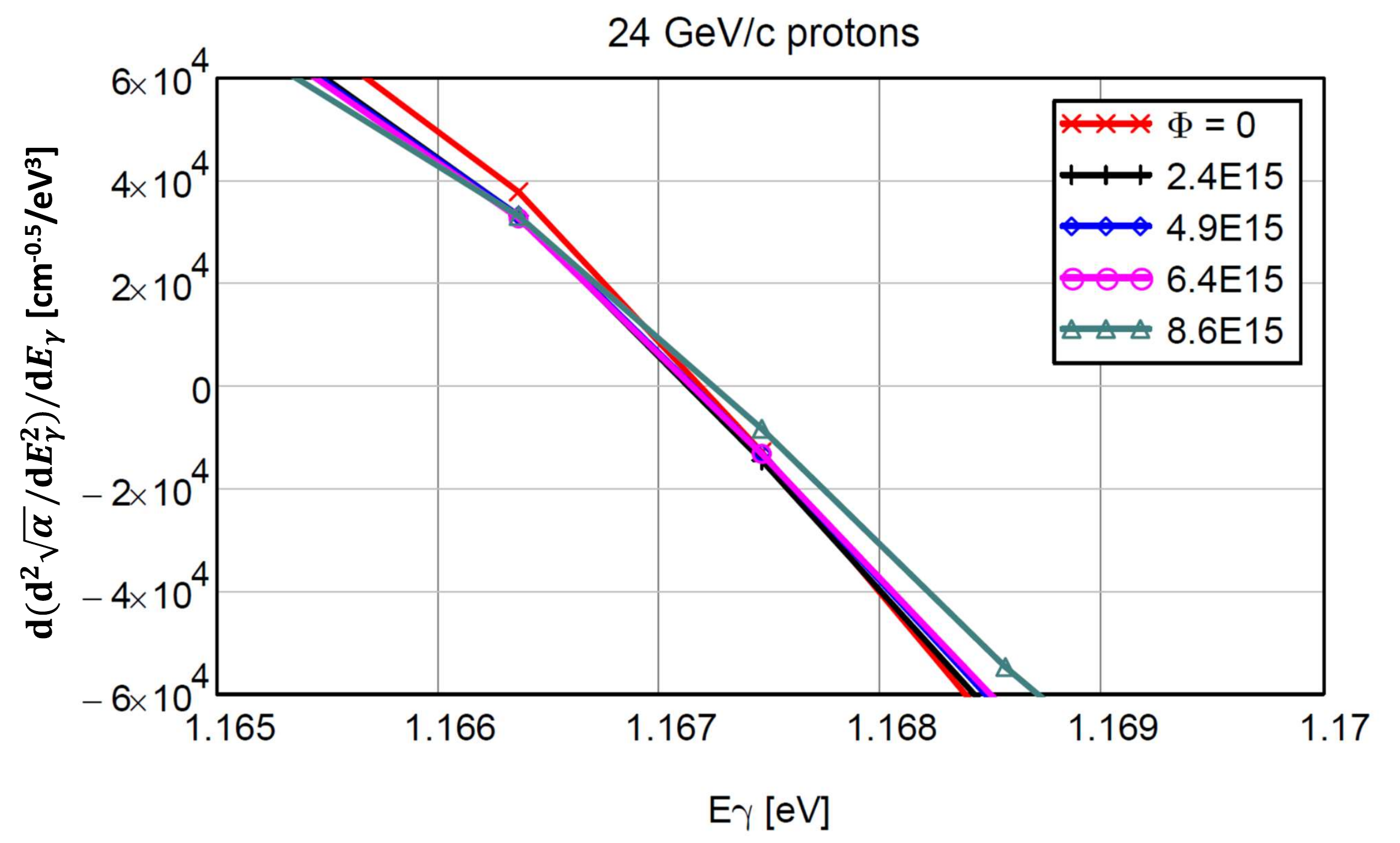}
    \caption{ }
    \label{fig:Figd3-protons-hi}
   \end{subfigure}%
   \caption{Determination of the peak positions using the zero of the first derivative of the smoothed second derivative, d$^2\sqrt{\alpha (E_\gamma)}/\mathrm{d}E_\gamma ^2$, shown in Fig.\,\ref{fig:Figd2-protons}.
   (a) $E_\gamma$ region of phonon absorption, and
   (b) $E_\gamma$ region of phonon emission. }
  \label{fig:Figd3-protons}
 \end{figure}

 The precise peak positions are determined by the zero crossing of the first derivative of d$^2\sqrt{\alpha}/\mathrm{d}E_\gamma ^2$.
 Examples in an expanded $E_\gamma$ scale for the data with proton irradiation are shown in Fig.\,\ref{fig:Figd3-protons}.
 For the phonon emission region, shown in Fig.\,\ref{fig:Figd3-protons-hi}, the zero crossings are within $\pm \, 0.1$\,meV.
 For the phonon absorption region, shown in Fig.\,\ref{fig:Figd3-protons-lo}, with the exception of the $\Phi _\mathit{eq} = 0$ data, the zero crossings are also within $\pm \, 0.1$\,meV.
 The agreement of the peak positions of d$^2\sqrt{\alpha (E_\gamma)}/\mathrm{d}E_\gamma ^2$ within a fraction of 1\,meV is remarkable.
 The results for the data with neutron irradiation are similar.
 From the values $E_\mathit{abs}$ for phonon absorption and $E_\mathit{em}$ for phonon emission, $E_\mathit{gap}$ and $E_\mathit{ph}$ are determined:

 \begin{equation}\label{eq:Egap}
   E_\mathit{gap} = 0.5 \cdot(E_\mathit{abs} + E_\mathit{em}) + E_\mathit{exc} \hspace{0.5cm} \mathrm{and} \hspace{0.5cm} E_\mathit{ph} = 0.5 \cdot(E_\mathit{em} - E_\mathit{abs}).
 \end{equation}

\begin{figure}[!ht]
   \centering
   \begin{subfigure}[a]{0.5\textwidth}
    \includegraphics[width=\textwidth]{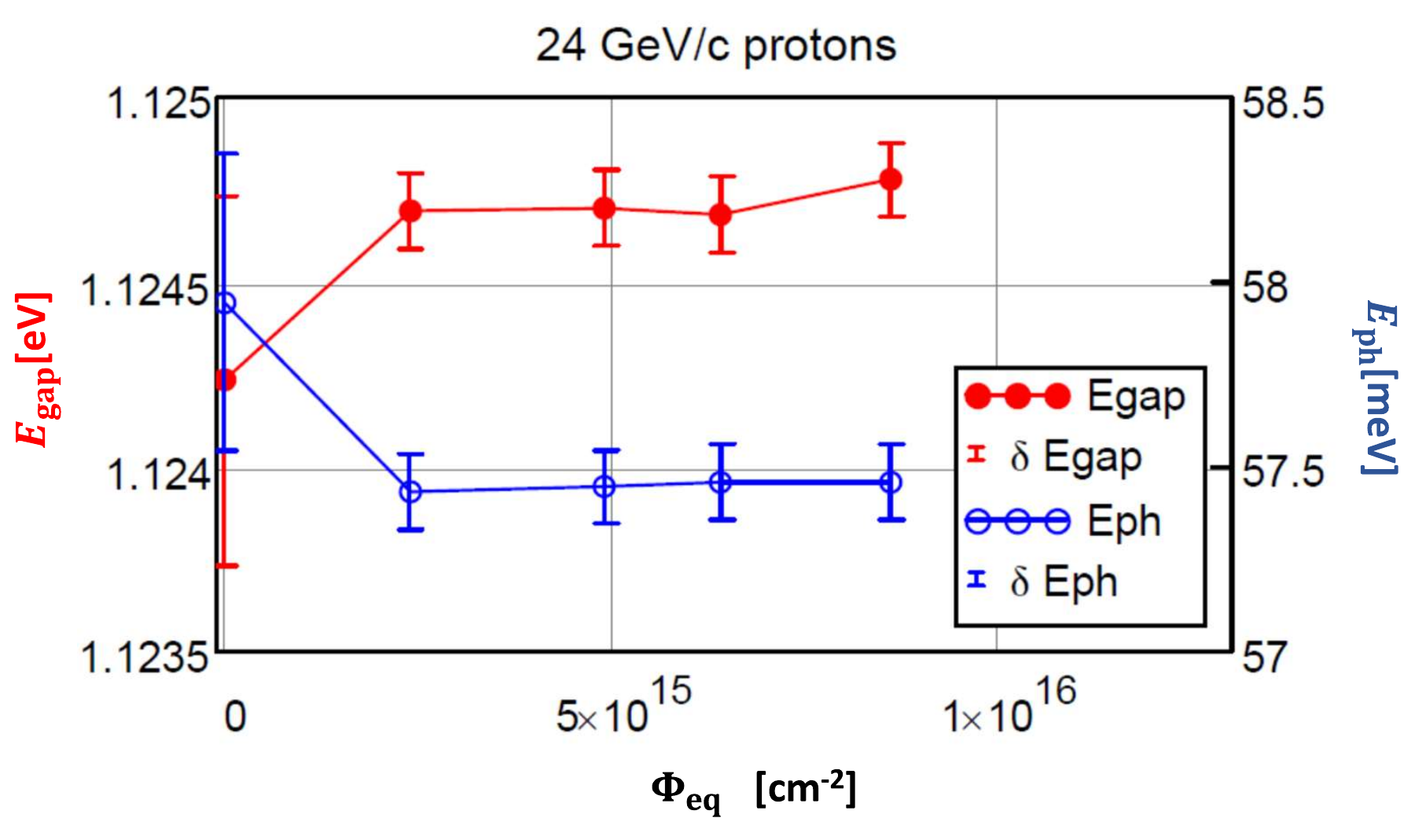}
    \caption{ }
    \label{fig:FigEgap-protons}
   \end{subfigure}%
    ~
   \begin{subfigure}[a]{0.5\textwidth}
    \includegraphics[width=\textwidth]{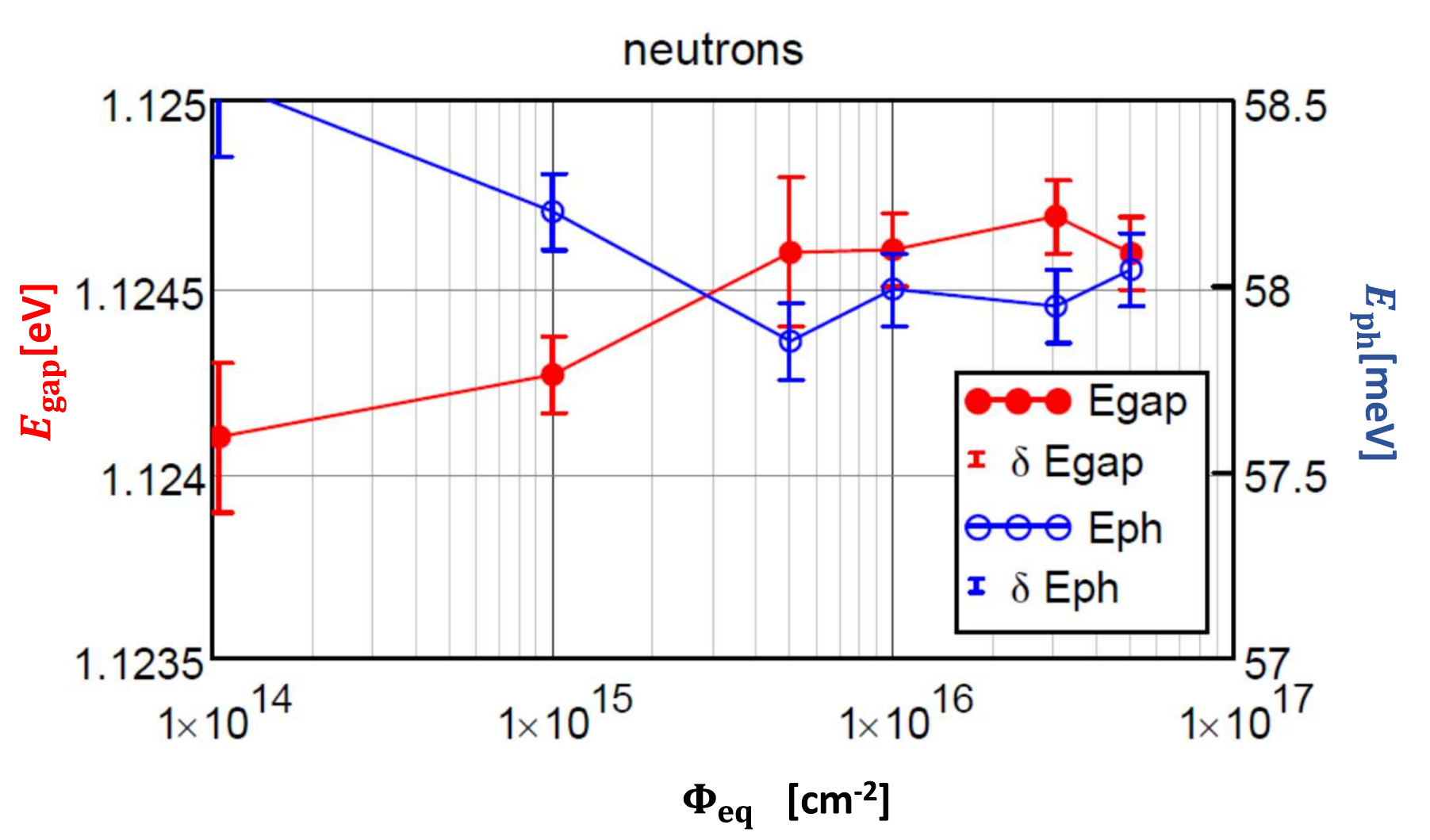}
    \caption{ }
    \label{fig:FigEgap-neutrons.pdf}
   \end{subfigure}%
   \caption{$E_\mathit{gap}$ and $E_\mathit{ph}$ as a function of $\Phi_\mathit{eq}$ for
   (a) proton, and (b) neutron irradiation.
   In (b) the $\Phi_\mathit{eq} = 0 $ result is shown at $10^{14}\,\mathrm{cm}^{-2}$.
   The error bars correspond to a change of the measured transmission by 1\,\%, which is an upper limit of the uncertainty of the transmission measurement.
   In addition the $E_\gamma $ scale has an uncertainty of about 1\,meV.}
  \label{fig:FigEgap-irr}
 \end{figure}

 The results for the $\Phi _\mathit{eq}$ dependence of $E_\mathit{gap}$ using $E_\mathit{exc} = 15$\,meV, and of $E_\mathit{ph}$ are shown in Fig.\,\ref{fig:FigEgap-irr}.
 It can be seen that within the fluctuations of the results, which are well below 1\,meV, neither $E_\mathit{gap}$ nor $E_\mathit{ph}$ depend on $\Phi _\mathit{eq}$.
 The systematic uncertainty due to the uncertainties of the transmission and wave-length scales of the spectral photometer is estimated to be 1\,meV, and the statistical uncertainty of the peak determination using the \emph{smoothed derivative} method 0.2\,meV.

 The values of $E_\mathit{gap} = (1.1245 \pm 0.0010)$\,eV and $E_\mathit{ph} = (58.0 \pm 1.0)$\,meV agree with the values from literature.

 The values for the neutron irradiation with $\Phi _\mathit{eq} = 1 \times 10^{17}\,\mathrm{cm}^{-2}$ are not shown, because noise prevents the determination of $E_\mathit{em}$.
 However, the value of $E_\mathit{abs} = 1.0516$\,eV agrees with the values at lower fluences.
 Finally, it is mentioned that the analysis of the 15\,min-isochronal annealing data between 80 and 210\,$^\circ$C shows that $E_\mathit{gap}$ and $E_\mathit{ph}$ do not change with annealing.

  To summarize this section:

  The \emph{smoothed derivative} method has been used to investigate if $E_\mathit{gap}$ and $E_\mathit{ph}$ of silicon change after irradiation with energetic protons and reactor neutrons.
  It is found that up to the investigated fluence of $1 \times 10^{17}\,\mathrm{cm}^{-2}$ possible changes of $E_\mathit{gap}$ and $E_\mathit{ph}$ are less than $\approx 0.5$\,meV.
  The values of $E_\mathit{gap}$ and $E_\mathit{ph}$ agree with the literature values at the 1\,meV level.

 \newpage

 \section{Band-gap energy in clusters}
  \label{sect:Egap-clu}

 Until now, the question, if the band-gap energy differs inside and outside of damage clusters, has not been answered.
 Problems are the large fluctuations of the cluster sizes and shapes, the poor knowledge of the cluster introduction rates and a probable dependence of $E_\mathit{gap}$ on the position within a cluster.
 In the following this question is addressed in a highly simplified approach:
 \begin{enumerate}
   \item The clusters are assumed to have an average volume $\langle V_\mathit{clu} \rangle$; it is estimated assuming that the cluster has the shape of a tube of length $l_\mathit{clu}$ and radius $r_\mathit{clu}$: $\langle V_\mathit{clu} \rangle = r_\mathit{clu}^2 \cdot \pi \cdot l_\mathit{clu}$.
       Inside of this volume the constant value $E_\mathit{gap}^\mathit{clu}$ is assumed for the energy band-gap, and outside $E_\mathit{gap} (\Phi = 0)$, the value measured for  non-irradiated silicon;
       $\Delta E = E_\mathit{gap}^\mathit{clu} - E_\mathit{gap}(0)$ is the difference of the band-gap energy inside and outside of a cluster.
   \item The cluster density is assumed to be $\rho_\mathit{clu} = \beta \cdot \Phi$, with the cluster-introduction rate $\beta$.
   \item The probability that the photon converts within a cluster is estimated using Poisson statistics: $p_\mathit{clu} = 1 - e^{-\rho_\mathit{clu} \cdot \langle V_\mathit{clu} \rangle}$.
   \item From the values of $E_\mathit{gap} (\Phi)$ determined in Sect.\,\ref{sect:Results}, $\Delta E = \left( E_\mathit{gap} (\Phi) - E_\mathit{gap} (0) \right) /  p_\mathit{clu}  $ is obtained.
 \end{enumerate}
 Finally, the formula used for estimating the energy-gap difference inside and outside of clusters is:

  \begin{equation}\label{eq:dE}
   \hspace{2.5cm} \Delta E = \frac{ E_\mathit{gap} (\Phi) - E_\mathit{gap} (0) } {1 - e^{- r_\mathit{clu}^2 \cdot \pi \cdot l_\mathit{clu}  \cdot \beta \cdot \Phi}}.
  \end{equation}

 As the density of cluster increases with $\Phi $, the sensitivity for $\Delta E$ increases with $\Phi$, and the $1 \times 10^{17} \mathrm{cm}^{-2} $ data are used in the following.
 From the data an upper limit of 0.5\,meV for the difference $ E_\mathit{gap} (1 \times 10^{17}\,\mathrm{cm}^{-2}) - E_\mathit{gap} (0)$ is obtained.
 For its determination, the energy threshold for phonon absorption, $E_\mathit{abs}$, had to be used as at this fluence noise does not allow to determine the threshold for phonon emission.
 It should be noted that, as also seen in Fig.\,\ref{fig:FigEgap-irr}, above  $\Phi = 1 \times 10^{15}\,\mathrm{cm}^{-2}$ the change of $E_\mathit{gap} (\Phi) $ is 0.1\,meV only, and the unexpected increase $E_\mathit{gap}$ at lower $\Phi $\,values is probably an artifact of the analysis procedure.

 A value $\beta \approx 0.15$ for the cluster introduction rate by 1\,MeV neutrons is extracted from Figs.\,5 and 7 of\,\cite{Huhtinen:2002}.
 For the cluster shapes, tubes of length $l_\mathit{clu}$ and radius $r_\mathit{clu}$ are assumed.
 From\,\cite{Huhtinen:2002} the value $\langle l_\mathit{clu} \rangle \approx  100$\,nm is estimated.
 The value $r_\mathit{clu} = 3.1$\,nm is obtained from $r_\mathit{clu} = a_\mathit{Bohr} \cdot \varepsilon _\mathit{Si} \cdot m_e/m_e^\ast$, with the Bohr radius of the hydrogen atom, $a_\mathit{Bohr} = 0.053$\,nm, the relative dielectric constant of silicon, $\varepsilon _\mathit{Si} = 11.7$, and the ratio of the effective to the free electron mass $ m_e^\ast / m_e = 0.2$ for silicon.
  The estimation of $r_\mathit{clu}$ is the back-of-the-envelope estimation of the effective radius of the electron cloud of donor atoms in silicon, resulting in $\langle V_\mathit{clu} \rangle = 3~000 \, \mathrm{nm}^3$.
 This value has a large uncertainty, which is also reflected by the numbers found in the literature, e.\,g. 3~800\,nm$^3$ in\,\cite{Lint:1972}, and 18~000\,nm$^3$ in\,\cite{Holmes:1970}.

 Inserting these numbers into Eq.\,\ref{eq:dE} results in values of $\Delta E = 11\, (2.3)$\,meV for $E_\mathit{gap} (1 \times 10^{17}\,\mathrm{cm}^{-2}) - E_\mathit{gap} (0)= 0.5\, (0.1)$\,meV, and a probability that the photon interacts inside a cluster of 4.4\,\%.

 These values can be compared to the results of~\cite{Donegani:2018}, where the change of the excitation energy, $\Delta E_a$, of damage states in clusters produced by 27\,MeV electrons has been investigated using TSC measurements.
 For the double vacancy $V_2$ the value $\Delta E_a = 7.6$\,meV was determined.
 It could also be shown that by annealing, $\Delta E_a$ approaches zero, which was taken as evidence that the clusters anneal.
 The change of $\Delta E_a$ has contributions from the electrostatic repulsion of charged states in the cluster, but can also come from a possible change of the energy-band gap.
 The values of $\Delta E_a$ are similar to the upper limits on $\Delta E$ obtained in this study, and thus can not answer the question how much a possible change of the band-gap energy contributes to $\Delta E_a$.

  \section{Conclusions}
  \label{sect:Conclusions}

 The main results of the paper are:
 \begin{enumerate}
   \item A method has been developed which allows to precisely locate kinks, i.\,e. sudden changes of the derivative as produced by thresholds, in measured spectra.
   \item The method is used to extract the band-gap energy, $E_\mathit{gap}$, and the energy of transverse optical phonons, $E_\mathit{ph}$, from optical transmission measurements of crystalline silicon irradiated by energetic protons and reactor neutrons to fluences up to $1 \times 10^{17}\,\mathrm{cm}^{-2}$.
   \item Within the 0.5\,meV accuracy, $E_\mathit{gap}$ and $E_\mathit{ph}$ do no change with irradiation fluence, and the values agree with results from literature.
 \end{enumerate}

 To the authors' knowledge, the method of determining the photon energies for phonon absorption, $E_\mathit{abs}$, and phonon emission, $E_\mathit{em}$,  from peaks in the second derivative of $\sqrt{\alpha (E_\gamma)}$ smoothed with a Gaussian kernel, has not been used so far.
 The accuracy achieved for $E_\mathit{abs}$ and $E_\mathit{em}$ is well below 1\,meV, and this method can be used to precisely determine the band-gap energy and the energy of the phonons required to excite an electron from the valence to the conduction band in indirect semi-conductors, like silicon.
 The results are hardly affected by the presence of light-absorption centers.
 For the measurements presented in this paper they are caused by the radiation damage of crystalline silicon by hadrons and dominate the absorption for the highest irradiation fluences investigated.

  For determining $\Delta E$, the difference of the band-gap energy inside and outside of clusters, the cluster introduction rate and the cluster shapes have to be known.
  Only crude estimates of these parameters are available, and the upper limit of $\approx 10$\,meV determined for $\Delta E$ has large uncertainties.


\section*{Acknowledgements}
 \label{sect:Acknowledgement}
  We thank Lukas Terkowski from the group of Roman Schnabel of the Institute of Laser Physics of Hamburg University for his help with the photo-spectrometer measurements, and Vladimir Cindro and Gregor Kramberger for the neutron irradiations performed at the TRIGA reactor of the Jozef Stefan Institute, Ljubljana.
  The  work was partially supported by the Deutsche Forschungsgemeinschaft (DFG, German Research Foundation) under Germany's Excellence Strategy -- EXC 2121 Quantum Universe -- 390833306, and by the "Partnership for Innovation, Education and Research (PIER)" between DESY and the University of Hamburg."

%
%



\end{document}